\documentstyle[epsf,twocolumn,prl,aps]{revtex}

\newcommand{\bra}{\langle}
\newcommand{\ket}{\rangle}

\begin{document}
\draft

 \twocolumn[\hsize\textwidth\columnwidth\hsize\csname @twocolumnfalse\endcsname

\title{Compact, orthogonal, and complete basis sets for solving
the Schrodinger equation }
\author{ Steven R.\ White}
\address{ 
Department of Physics and Astronomy,
University of California,
Irvine, CA 92697
}
\date{\today}
\maketitle
\begin{abstract}
\noindent 
We present a new type of basis set which is local, compact, and
orthogonal.  The basis functions, called orthlets, are centered at
the sites of a lattice and are specifically
adapted to represent the system being studied. The adaptability
includes the ability to have singular behavior within an
orthlet, allowing a single orthlet to represent a function in
the vicinity of a singularity. 

\end{abstract}
\pacs{PACS Numbers: }

 ]

Most modern numerical solution techniques to the Schrodinger
equation begin with the introduction of a basis set, thereby 
making an infinite Hilbert space finite. Because there are a
number of incompatible qualities one would like in a
basis set, a wide variety of basis sets are in use, each of
which is better behaved according to some set of criteria.
Among the desirable qualities are orthogonality, locality, 
compactness (i.e. compact support), 
the ability to represent space uniformly, the
ability to represent singular regions with higher resolution,
the ability to incorporate prior knowledge about singular
regions, the ability to ignore empty regions,
and the availability of specialized efficient algorithms 
(such as the fast fourier transforms (FFTs) or wavelet
transforms) for doing integrals and solving differential
equations.

For example, in electronic structure calculations for solids
using density functional methods, plane waves are widely
and successfully used\cite{payne}.
These are orthogonal, have uniform resolution, and the FFT
allows rapid switching between real and fourier space.
Pseudopotentials are normally used to represent atomic cores.
However, for more accurate treatment of the cores, for nonperiodic
systems (including molecules and surfaces), and for more
accurate treatment of interatomic correlations, the
plane wave basis is inconvenient.

In quantum chemistry, the standard choice for basis functions 
is the product of a radial
function centered on an atom times a cartesian or spherical harmonic\cite{qc}.
Because the radial functions which solve the Hartree Fock equations for
atoms are known, remarkably small numbers of basis
functions are needed---often only about 
twice as many basis functions as there are electrons.
The nonorthogonality of the basis is easily dealt with in Hartree Fock.
The major drawbacks relate to scaling to large
systems and to high accuracy. The number of two-electron
integrals needed to represent the inter-electron Coulomb
interaction scales as $N^4$, where $N$ is the number of basis
functions.  Moreover, the orthogonalization required for most
treatments of correlations beyond Hartree Fock destroys the
approximate locality of the functions; consequently, 
computation time typically scales as $N^6$ or worse. 

Wavelet bases are another potentially attractive
alternative\cite{arias}.
These nonorthogonal bases allow for
widely varying resolution to represent both cores and valence
electrons. However, hundreds or thousands of wavelets on various
length scales may be needed to represent a second row atomic
core, compared to perhaps a dozen
of the radial basis functions used in quantum chemistry,

In this letter, we propose a new type of basis set which is
orthogonal, very localized and compact, which allows variable
resolution, and which allows prior knowledge about singularities
to be incorporated into the basis, while keeping the number of
basis functions to a minimum. Our approach is most closely
related to the finite element basis using orthogonal shape
functions developed by White, Wilkins, and Teter (WWT)\cite{wwt}.
The major problem with the approach of WWT was the difficulty
in obtaining adequate resolution for the cores. Our new approach
overcomes that difficulty.
Although these bases were developed with electronic structure
calculations in mind, we expect them to be useful in a
variety of other contexts as well.

Consider a set of localized {\it shape }
functions $\phi_i(\vec r)$, and a lattice
$\{\vec R_j\}$. We can generate a set of functions for each
lattice site by translation, $\phi_{ij} = \phi_i(\vec r - \vec R_j)$.
Let the functions have the following 
properties: 1) the set of functions $\phi_i(\vec r)$ is
orthonormal; 2) each function is also orthogonal to the
functions on all other lattice sites; and 3) the total set
of functions on all lattice sites is {\it complete}.
(Wannier functions also have these properties.)
We define a projection operator for site $j$ by
\begin{equation}
P_j = \sum_i | \phi_{ij} \ket \bra\phi_{ij} | .
\end{equation}
In coordinate notation this operator is 
\begin{equation}
P_j(\vec r,\vec r') = \sum_i \phi_i(\vec r - \vec R_j)
\phi_i(\vec r' - \vec R_j) ,
\end{equation}
where the $\phi_i$ are assumed real.
Completeness implies that $\sum_j P_j = 1$. Now consider the
application of $P_j$ on an arbitrary function $f(\vec r)$:
$f_j(\vec r) = P_j f(\vec r)$. Then 
\begin{equation}
f(\vec r) = \sum_j f_j(\vec r) ,
\end{equation}
and
\begin{equation}
\bra f_j(\vec r) | f_{j'}(\vec r) \ket = 0, \ \ \ \ \ j \ne j' .
\end{equation}
We will call functions such as $f_j(\vec r)$, which are
local, orthogonal, and specifically adapted to a function or to a set of
functions, {\it orthlets}. 
The basis formed by the orthlets $f_j(\vec r)$
is in a sense a perfect basis for representing $f(\vec r)$: it
is orthogonal, represents the function exactly, and has the
minimum number of functions given the scale set by the lattice
spacing.  We describe below how to form an orthlet basis describing an
arbitrary {\it set}  of functions $f^\alpha(\vec r)$ to a specified
accuracy.

In order for the orthlets to be useful, the shape functions
$\phi_i(\vec r)$ should be smooth and local, and preferably compact.
In two or more dimensions, shape functions can be written as
products of one dimensional shape functions\cite{wwt}, so that 
we need only consider the 1D case.
WWT developed a set of four shape functions with continuous derivatives up
to third order, which were able to represent exactly polynomials
up to third order\cite{wwt}. These shape functions were compact, with 
a total width of two, where we assume a lattice spacing of 
unity henceforth. The compactness means that orthogonality must
only be specifically arranged for nearest neighbor functions.
Here we give six orthogonal shape functions which are smooth, i.e. all
derivatives are continuous, which also are compact with width
two, and which are able to represent polynomials exactly up to
order five. We believe these shape functions
are sufficiently complete for most uses, because we give
alternative ways of generating the orthlets in the vicinity of
singularities and also for changing lattice spacings. The
smoothness allows additional functions to be added to increase
completeness without adjusting the functions one already has.

All shape functions $S_n(x)$ are defined for $x \ge 0$; $S_n(x)$
is even (odd) if $n$ is.
We construct $S_0(x)$ to represent a constant
function exactly. We first define a smooth ``splicing function'' 
$p(x)$ which divides a function to be fit to into pieces. We require that
$p(x) = 0$ for $|x| \ge 1$, $p(x) = p(-x)$, and that 
\begin{equation}
p(x) + p(x-1) = 1 \ \ \ \ \ 0 \le x \le 1 .
\end{equation}
We choose the function
\begin{equation}
p(x) = \frac{1}{2} - \frac{1}{2} \tanh(\frac{3}{2} \ 
\frac{x-1/2}{[x(1-x)]^{1/2}})	\ \ \ \ \ 0 < x < 1 .
\end{equation}
This bell-shaped function has essential singularities at $x = 0, \pm 1$,
allowing it to have compact support yet be smooth.

The shape function is obtained by multiplying the function to
fit to, in this case unity, by $p(x)$, and then adding a smooth 
oscillating function $o(x)$ to induce orthogonality. The fit
will not be spoiled if
\begin{equation}
o(x) + o(x-1) = 0 \ \ \ \ \ 0 \le x \le 1 .
\end{equation}
We choose
\begin{equation}
o(x) = \sum_{m=1}^M a_{0m} o_m(x)
\end{equation}
where
\begin{equation}
o_m(x) = p(2x-1) \sin(2 m \pi (x-1/2)).
\end{equation}
For the first shape function, $S_0(x)$, is suffices to take $M=1$, with
$a_{01} = -0.507021142747521$:
\begin{equation}
S_0(x) = p(x) + a_{01} o_1(x) \ \ \ \ \ 0 \le x \le 1 .
\end{equation}
Note that $S_0(x) + S_0(x-1) = 1$ for $0 \le x \le 1$, 
so that $S_0(x)$ is already normalized. We show $S_0(x)$ in
Fig. 1(b).

The second shape function is obtained similarly, by requiring an
exact fit to the function $x$. First, we attempt to fit the
function $x$ using only $S_0(x)$. We then make $S_1(x)$ out
of the error or residual of this fit,
\begin{equation}
r_1(x) = x - S_0(x-1) \ \ \ \ \ 0 \le x \le 1 .
\end{equation}
In this case, the extra orthogonalizing functions must be made
out of cosine functions rather than sine functions, because the
function itself is odd. For $0 \le x \le 1$
\begin{equation}
S_1(x) = N_1 \left( p(x)r_1(x) + \sum_{m=1}^2 a_{1m} e_m(x) 
\right) ,
\end{equation}
with $S_1(-x) = -S_1(x)$, 
\begin{equation}
e_m(x) = p(2x-1) \cos((2 m - 1) \pi (x-1/2)),
\end{equation}
$a_{11} = 0.132403793351197$, 
$a_{12} = -0.048844623781880$, and
$N_1 = 3.78750743638139$.

\begin{figure}[ht]
\epsfxsize=2.5 in\centerline{\epsffile{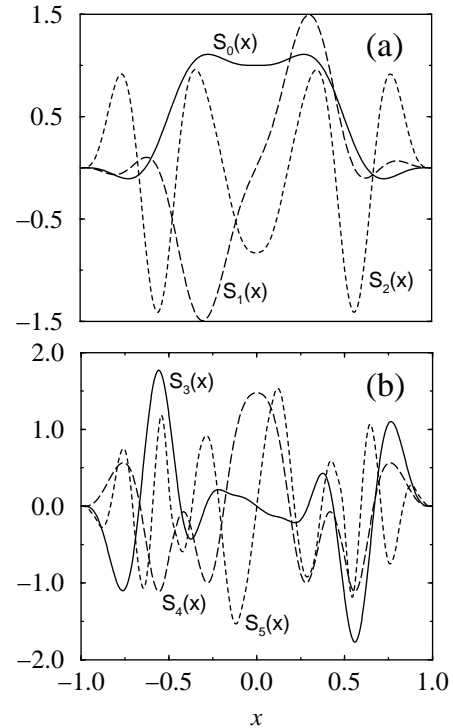}}
\caption{(a) and (b) Shape functions $S_n(x)$.
}
\end{figure}

To determine the coefficients $a_{1m}$, and later coefficients for higher
order shape functions, two different linear combinations of
the $e_m(x)$ or $o_m(x)$ were found which induced orthogonality to 
all the lower order shape functions centered at $x=1$. These
two different combinations were then combined in order to induce
orthogonality of $S_n(x)$ and $S_n(x-1)$. This last procedure
required solving a quadratic equation, which might not have real
roots. If there were no real roots, more orthogonalizing
functions were included. A solution in this case required
finding both positive and negative eigenvalues of a symmetric
matrix, which also sometimes did not occur. There does not
appear to be any guarantee of a real solution; in fact, we did
not find a satisfactory $S_6(x)$ able to fit $x^6$ exactly.

The next shape function is given by
\begin{eqnarray}
S_2(x) = N_2 [  p(x)(x^2 - c_{20} S_0(x)
- d_{20} S_0(x-1) \\ \nonumber  - d_{21} S_1(x-1))
+ a_{21} o_1(x) + a_{22} o_2(x) + a_{23} o_3(x)
],
\end{eqnarray}
$c_{20} = 0.0697096675548214$,
$d_{20} = 1.0697096675548214$,
$d_{21} = 0.528051768503122$,
$a_{21} = 0.088401702549656$,
$a_{22} = -0.126644764032427$,
$a_{23} = -0.025357986069321$, and
$N_2 = 11.9312518524753$.
Expressions for $S_3$ to $S_5$ will be presented
elsewhere\cite{website}.

In Fig. 2, we use these shape functions to generate
an orthlet basis for a function with a slope
discontinuity. For the lattice sites away from the singularity,
the overlap integral of the function with each shape function
was computed numerically.
Resumming the shape functions on a site with these integrals as
coefficients gives the orthlet. In the case of
the orthlet at $x=0$, where the slope discontinuity is, an
expansion would converge too slowly. Instead, the orthlet was
obtained by subtracting from $f(x)$ the orthlets on the adjacent
sites. 

\begin{figure}[ht]
\epsfxsize=3.0 in\centerline{\epsffile{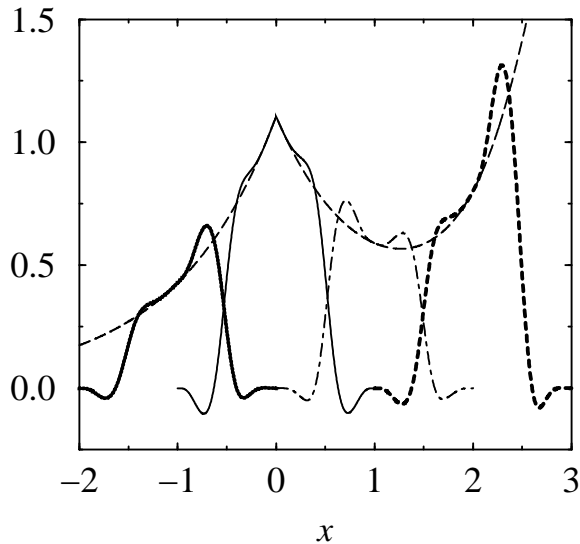}}
\caption{Orthlets used to fit the function
$f(x)=\exp(-|x|)+1/((x-3)^2 + 1/2)$. The sum of these
four orthlets is equal to $f(x)$ within the region $-1 \le x \le 2$.
}
\end{figure}

If the shape functions describe the
function perfectly in the region away from the singularity, then
the subtracted function will be identically zero for $|x| \ge 1$.
If the fit is not perfect but very good, the function can be set to zero
for $|x| \ge 1$. However, a small discontinuity and a
small lack of orthogonality can result from this procedure. 
In this example,  the discontinuities were less than $10^{-5}$ and
were ignored, along with a small nonorthogonality.
In order to ensure perfect continuity and orthogonality in this example,
we could multiply the subtracted function by a smooth windowing function which
is unity for most of the interval $-1 \le x \le 1$, 
and is zero for $|x| \ge 1$.
Then the resulting function can be explicitly orthogonalized in
a Gram-Schmidt (GS) procedure to the neighboring shape functions.
Because of the orthogonalization, 
the resulting function would extend from $-2 \le x \le 2$.

Another procedure for dealing with singularities is to add
a set of localized functions near the singularity, which are
chosen for their ability to represent the singularity.
Then an explicit GS orthogonalization
procedure mixes these functions with the shape functions
that overlap them. This GS procedure does not
destroy overall locality: all shape functions not directly overlapping the
functions added would still automatically be
orthogonal to the new set of functions. 
Orthlets are then formed as linear combinations of the shape
functions and the added singularity functions.
In three dimensions, we form shape functions as cartesian products of
the 1D shape functions, 
$S_{\vec n}(\vec r) = S_{n_x}(x) S_{n_y}(y) S_{n_z}(z)$.
In preliminary
calculations to represent the hydrogen atom ground state,
we have found that adding a set of narrow gaussians, multiplied
by a windowing function similar to $p(x)$, is very convenient
and effective for representing the cusp singularity at the nucleus. 
Part of the convenience is that 3D gaussians are products of 1D
gaussians, like the 3D shape functions.
These results will be presented elsewhere.

There are several ways to change the basic lattice spacing
in different regions of space. One simple approach is to
let the finer and coarser grids overlap, so that
completeness is ensured,  and
then apply a GS procedure to the overlap region,
automatically generating local functions which connect the two
regions. However, the ability to generate orthlets with 
singularities means that changing the lattice spacing is usually
not necessary.

Now, suppose
one wishes to generate orthlets to represent a set of functions 
$\{f^\alpha(\vec r)\}$. For example, one might want to build an
orthlet basis which is able to represent a standard radial basis set
from quantum chemistry, since these basis sets are known to
represent Hartree Fock orbitals well.
Note that the $f_\alpha$ need not be orthogonal, but the orthlet
basis generated from them is. The orthlet basis will automatically
have additional degrees of freedom allowing improved treatment of
correlations.
For simplicity, we will let $S_{\vec n}(\vec r)$ represent
both shape functions and any extra singularity basis functions.
To adapt the basis to represent the $f_\alpha$, we apply the
procedure in the density matrix renormalization group for
targetting more than one state\cite{dmrg}.
For each lattice site $j$, we find the coefficients
\begin{equation}
c_{\alpha \vec n} = \bra S^j_{\vec n}(\vec r) | f^\alpha(\vec r)\ket.
\end{equation}
We now form the positive semidefinite density matrix
\begin{equation}
\rho^j_{\vec n' \vec n} = \sum_\alpha c_{\alpha \vec n'} c_{\alpha \vec n}.
\end{equation}
Note that positive weighting
factors $a_\alpha$ can also be included in the sum if some of
the functions are considered more important than others.
The eigenvectors of $\rho^j$, 
$v^\beta_{\vec n}$, define orthlet functions 
$v^\beta(\vec r) = \sum_{\vec n} v^\beta_{\vec n} S^j_{\vec n}(\vec r)$ which
optimally represent the functions $\{f^\alpha(\vec r)\}$ in the
basis for site $j$. Each density matrix eigenvalue
$w_\beta$ gives
the weight associated with that orthlet in representing the
$f_\alpha$. By choosing a cut off weight, and retaining all
orthlets with weight greater than the cut off, one obtains a
systematically improvable basis for the $f_\alpha$. 
In particular, one can show that
\begin{equation}
w_\beta = \sum_\alpha  (\bra v^\beta | f^\alpha \ket)^2 .
\end{equation}
Thus, if a density matrix eigenvalue $w_\beta$ is very small,
then none of the $f^\alpha$ have significant overlap with the
corresponding function $v^\beta$, and $v^\beta$ need not be
included in the basis. 
The basis set becomes exact if the number of orthlets kept per
site is equal to the number of functions $f_\alpha$.
If only one function is in the set
$f_\alpha$, then there is only one nonzero eigenvalue and the
orthlet is simply the normalized projected function $P_i
f_\alpha$.  

As an example of this procedure, we generate a general set of
orthlets to use as a basis set for the tails of wavefunctions. The
ordinary shape functions are too localized to represent an
exponentially decaying tail efficiently. The orthlets
we generate here extend quite far in one direction, and so are
very useful to use as replacements for the shape functions for the edge
sites of the lattice, allowing fewer sites to be used. 
The functions are constructed
by using as $f_\alpha$ a set of 13 gaussians with widths ranging from
1 to 4, each centered at the origin, constructing and diagonalizaing
the density matrix.  The basis for an orthlet
includes more than one site here: we include all 78 shape
functions on sites 3 to 15. The resulting orthlets are able to
represent any linear combination of the 13 gaussians with
good accuracy.  The three most important functions, 
along with the density
matrix eigenvalues $w_\beta$, are shown in Fig. 3. One can see
the the density matrix eigenvalues decay very rapidly, so that
only a few of these orthlets are needed. 

In electronic structure calculations, one would use a lattice spacing
appropriate for the valence electrons, say 0.3-1.0 angstroms.
Cores would be treated using
orthlets derived from localized cusp functions, which would be
tied to each nucleus. 

\begin{figure}[ht]
\epsfxsize=2.5 in\centerline{\epsffile{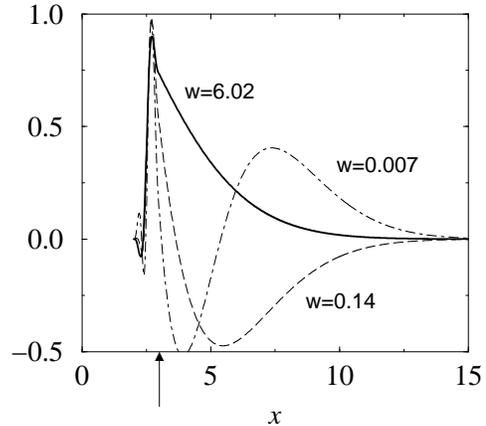}}
\caption{Orthlets which can be used to represent tails of
wavefunctions. The orthlets are adapted to a set of 13 
gaussians centered at 0 with a variety of widths. 
The density matrix eigenvalue $w$ 
of each orthlet is given. The arrow indicates the location of
the first site (3) included in the basis for the orthlets. The
functions are orthogonal to all shape functions to the left of
that site.
}
\end{figure}

Lattice sites near
cores would have a dozen or more orthlets; in other areas,
we expect only a few might be needed, making perhaps 100-300
basis functions per atom. Although this is
an order of magnitude more than with the radial functions used
in quantum chemistry, it is perhaps an order of magnitude less
than a wavelet basis, and the orthlets would be orthogonal and
compact.  The orthlets appear to be very convenient for the development of
$o(N)$ algorithms for density functional calculations; for
example, sparse matrix methods coupled with the multigrid
algorithm might be used to solve the Poisson equation. 
Orthlets also appear well adapted to multipole
representations of the electron-electron interaction, which
can also be used in $o(N)$ algorithms.

We acknowledge support from the NSF under 
Grant No. DMR-9870930.

\end{document}